\documentclass{iau_FM}
\usepackage{graphicx}

\title[The Role of Magnetic Fields in The Evolution of Galaxies] 
{The Role of Magnetic Fields in The Evolution of Galaxies}

\author[Fatemeh S. Tabatabaei]   
{Fatemeh S. Tabatabaei
 }
 
\affiliation{Instituto de Astrof\'isica de Canarias, V\'ia L\'actea S/N, E-38205 La Laguna, Spain\\ Departamento de Astrof\'{i}sica, Universidad de La Laguna, E-38206 La Laguna, Spain\\School of Astronomy, Institute for Research in Fundamental Sciences, 19395-5531 Tehran, Iran\\ email: {\tt ftaba@ipm.ir}}

\pubyear{2018}
\jname{Astronomy in Focus} 
\editors{Teresa Lago, ed.}
\begin{document}

\maketitle

\begin{abstract}
Magnetic fields constitute an energetic component of the interstellar medium in galaxies and, hence, can affect the formation of galactic structures. Sensitive resolved radio continuum observations together with statistical studies in galaxy samples are performed to investigate the origin and the impact of the  magnetic fields. The JVLA cloud-scale survey of M33 unveils strong tangled magnetic field along the spiral arms and in the galaxy center indicating amplification due to compression and local shear in molecular clouds. Studying a sample of non-cluster, non-interacting galaxies, we find that the large-scale ordered magnetic field scales with the rotation speed of galaxies. The total and disordered magnetic fields scales with the star formation rate in normal star forming galaxies. On the other hand, a strong magnetic field in the center of NGC\,1097--a massive galaxy undergoing star formation quenching-- is found to be responsible for decelerating its massive star formation. Putting these results together, it is deduced that 1) the Universe was highly magnetized short after the peak of the massive star birth at about 1\,Gyr after the Big Bang and 2) the strong magnetic field has possibly acted as back reaction after this time, quenching the massive star formation and stimulating the formation of low-mass stars in massive galaxies.

\end{abstract}

\firstsection 
\section{Introduction}
Cosmic magnetic fields have been present since the very early Universe created through mechanisms such as plasma instabilities and cosmic batteries (e.g., \cite[Subramanian 2016]{sub16}). Theoretical studies show that, after the epoch of reionization, these weak seed fields ($\lesssim 10^{-16}$\,G) were amplified to stronger fields of $\gtrsim \mu$G through various processes such as compression and shear motions (\cite[Beck et al. 2013]{beck13}) or small-scale dynamos (e.g., \cite[Schober et al. 2013]{scho13}) on time-scales of few to hundreds of $Myr$. The amplification due to compression and shear occurs because the magnetic field lines are frozen into a fluid and due to the magnetic flux conservation law. The dynamo action, that is based on the energy conservation, converts kinetic energy from turbulence to magnetic energy. Theoretically, both mechanisms can occur in galaxies, and particularly in star forming regions where both supernova-driven turbulence and bulk motions of dense gas are present, however, a general picture of the origin of the field amplification in the interstellar medium (ISM) is missing and awaits sensitive and high-resolution observations in statistically meaningful galaxy samples. 

Full polarization radio continuum observations make it possible to measure and map the magnetic field strength and structure in galaxies (see the review by \cite[Beck 2015]{beck15}). However, the radio continuum emission needs to be corrected first for contamination by the thermal free-free emission. This is particularly important in high-resolution studies of galaxies even at low frequencies. The dust-unbiased {\it Thermal Radio Template (TRT)} methods, developed by \cite[Tabatabaei et al. (2007)]{tab7}, \cite[Tabatabaei et al. (2013)]{tab13}, and \cite[Tabatabaei et al. (2018)]{tab18}, map the thermal and the nonthermal synchrotron emission in galaxies without any assumption about the synchrotron spectrum. These methods are sensitive to the diffuse ISM structures and hence ideal to study the role of the magnetic fields in the energy balance and structure formation in different locations in a galaxy. 
In addition to the high-resolution studies, surveys in galaxy samples are vital to probe any dependence on galactic properties such as mass, morphology, dynamics, and star formation rate shedding light on the origin and amplification mechanisms of the magnetic fields.  Our recent observations in nearby galaxies and their implications for the evolution of galaxies are described as follows.   
\vspace*{-0.5 cm}

\begin{figure}[b]
\begin{center}
\includegraphics[width=4.5in]{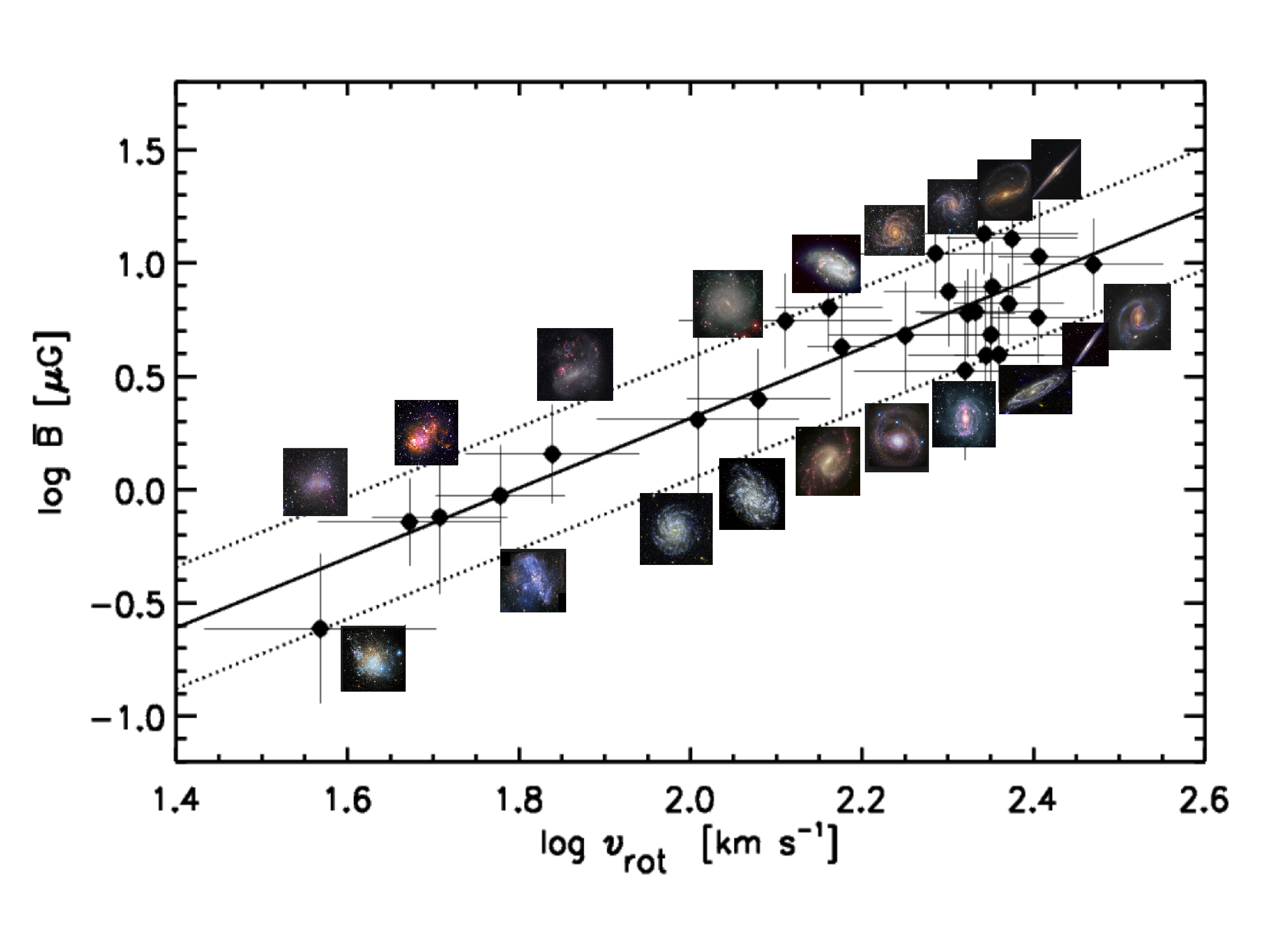} 
 \vspace*{-0.5 cm}
\caption{The large-scale magnetic field strength $\overline{B}$ vs. the rotational speed of a sample of nearby non-cluster/non-interacting galaxies with available full polarization radio continuum data (\cite[Tabatabaei et al. 2016]{tab16}). Faster rotating and/or more massive galaxies have stronger ordered magnetic fields (with scales of $\gtrsim$1\,kpc). This figure shows the case of the equipartition between the magnetic field and cosmic rays, $\overline{B} \propto v_{\rm rot}^{1.5}$. A linear correlation holds assuming that the cosmic ray number is proportional to the star formation rate (not shown). }
\end{center}
\end{figure}

\section{Large-scale magnetic field and galactic dynamics}
The observed polarized radio continuum emission from spiral galaxies exhibits a large-scale ordered pattern which is often linked to the mean field $\alpha-\Omega$ dynamo (e.g., \cite[Beck 2015]{beck15}) if the observed pitch angles agree with theoretical estimates (\cite[Shukurov 2004]{shuk}). However, a global picture connecting the strength of the large-scale magnetic field to dynamics of galaxies have been awaiting further observations to build a statistically meaningful sample. Defining this sample, it is also necessary to take into account the galactic environment as galaxy dynamics can be influenced by interactions and mergers.  Putting together the available data of non-interacting, non-cluster galaxies with known large-scale magnetic fields, \cite[Tabatabaei et al. (2016)]{tab16} find a tight correlation between the strength of the integrated polarized flux density and the rotation speed $v_{\rm rot}$ of galaxies. This translates to an almost linear correlation between the large-scale magnetic field $\overline{B}$ and $v_{\rm rot}$ assuming that the number of cosmic-ray electrons is proportional to the star formation rate, and a super-linear relation, $\overline{B} \propto v_{\rm rot}^{1.5}$, assuming equipartition between magnetic fields and cosmic rays (Fig. 1). It is important to note that this correlation is not due to the mean field $\alpha-\Omega$ dynamo, as no correlation holds between $\overline{B}$ and the differential rotation (or angular velocity $\Omega$ for flat rotation curves). Instead, it shows a coupling between the ordered field and the dynamical mass of galaxies and indicates that gas compression and/or shear is an important mechanism ordering the large-scale ($\gtrsim$1\,kpc) magnetic field in galaxies. \vspace*{-0.5 cm}
\section{Origin of small-scale magnetic field}
Only a minor fraction of the observed synchrotron emission from spiral galaxies is polarized (usually $<20\%$ at 4.8\,GHz) meaning that a major part of the emission is due to a field that is varying on scales smaller than observation beam size. This field can be isotropic random and/or non-isotropic tangled field. The maps of the disordered field strength in star forming galaxies shows higher values in massive star forming regions (e.g., $B_{\rm dis} \propto \Sigma_{\rm SFR}^{0.16}$, \cite[Tabatabaei et al. 2013]{tab13}) indicating possible act of the small-scale dynamo, although the supernovae-driven dynamo predicts a somehow higher power-law index of 0.3 (\cite[Gressel et al. 2008]{gres}). This theoretical index matches better with the global studies of the {\it KINGFISHER} sample ($B \propto \Sigma_{\rm SFR}^{0.3}$, \cite[Tabatabaei et al. 2017]{tab17}), the {\it SINGS} sample (\cite[Heesen et al. 2014]{Hee}), and a number of dwarf galaxies (\cite[Chy\.zy et al. 2011]{chyz}).

The JVLA full-polarization observations of M33 at 6\,GHz (5-7\,GHz, Tabatabaei et al. in prep.) unveils a totally different view of the magnetized ISM at 9'' resolution (36\,pc, scale of giant molecular clouds). The linearly polarized emission vectors nicely point towards the molecular clouds in star forming complexes and in the spiral arms exhibiting the tangled magnetic field. This field was previously undetected due to depolarization at the large beams of $\gtrsim$120'' (\cite[Tabatabaei et al. 2008]{tab8}). This also shows the importance of the compression/shear in ordering and amplification of the magnetic field on cloud scales.  \vspace*{-0.5 cm}

\begin{figure}
\begin{center}
\resizebox{\hsize}{!}{\includegraphics*{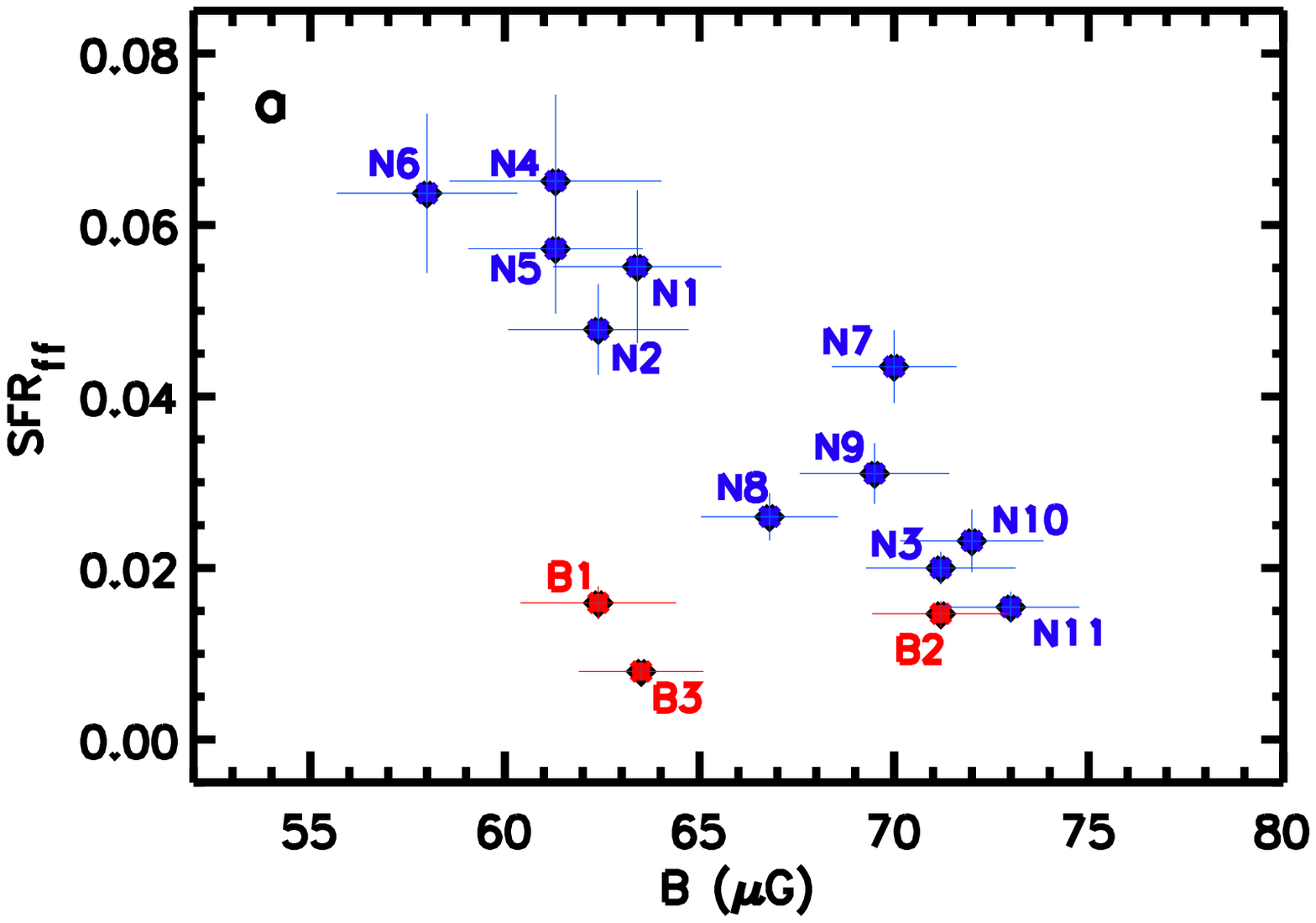}\includegraphics*{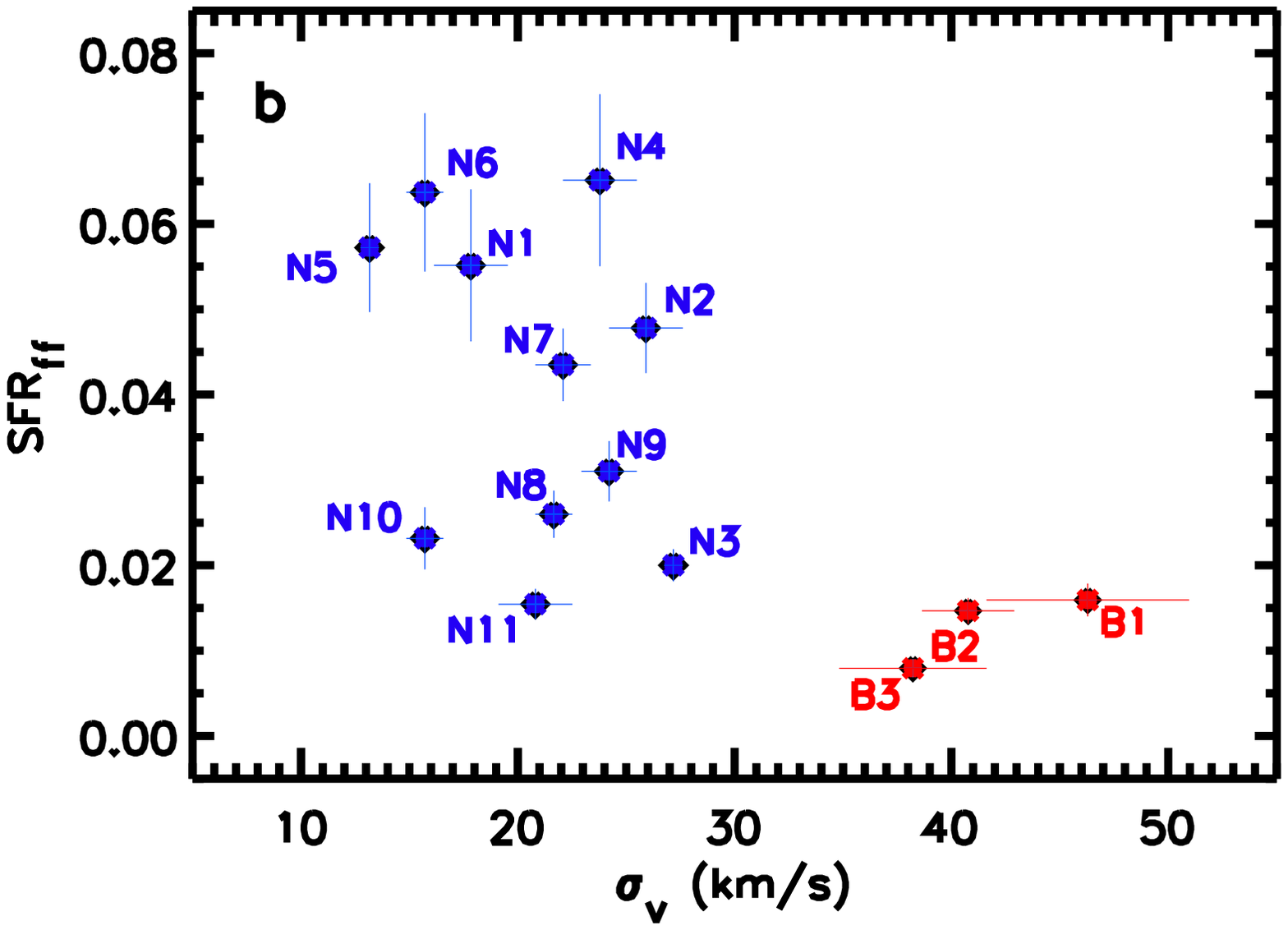}}
\caption[]{The massive star formation rate per free-fall, SFR$_{\rm ff}$, of the GMAs decreases with the magnetic field strength B ({\it a}) while it is uncorrelated with the turbulent velocity $\sigma_v$ ({\it b}).  The blue and red points show the narrow-  and broad-line GMAs, respectively. Strong non-circular motions/shocks in the broad-line GMAs can act as additional cause of the low SFR$_{\rm ff}$ in these clouds (\cite[Tabatabaei et al. 2018]{tab18}).}
\label{fig:B}
\end{center}
\end{figure}
\section{Magnetic feedback controlling ISM structure and star formation}
Investigating the ISM energy balance in the central kpc region of the quenching galaxy NGC\,1097, \cite[Tabatabaei et al. (2018)]{tab18} find a major role of the magnetic field and cosmic rays in controlling the molecular clouds and star formation. Most of the giant molecular cloud associations (GMA) are magnetically supported against gravitational collapse and the star formation rate per free-fall drops with the magnetic field strength, while it is uncorrelated with the turbulence (Fig. 2). This explains the deviations from the star formation laws observed earlier by \cite[Hsieh et al. (2013)]{Hsieh}. \vspace*{-0.5 cm} 
\section{Implications: magnetic field \& quenching of galaxies}
The dependency of the magnetic field strength on the star formation rate has an important consequence for the evolution of galaxies. As the cosmic star formation rate peaks at about 1\,Gyr after the Big Bang, the magnetic field should also reach its maximum strength through compression and/or small-scale dynamo in a relatively short time scale. Considering that galaxies start to evolve after this time ($t\gtrsim$1\,Gyr), it is then plausible that the magnetic field plays a role in this evolution. Recent discovery by \cite[Tabatabaei et al. (2018)]{tab18}, shows that the strong magnetic field can decelerate massive star formation and can help the formation of low-mass stars. Hence, magnetic field has possibly acted as a back reaction after the peak of the star formation rate in the Universe, quenching massive star formation and stimulating the formation of big bulges of low-mass stars observed in quenched massive galaxies (e.g., \cite[Bell 2008]{bell}). 
%
%
%


\end{document}